\documentclass[a4paper]{jpconf}
\usepackage{graphicx}

\newcommand{\R}{\mathcal{R}}
\def\be{\begin{equation}}
\def\ee{\end{equation}}
\def\bea{\begin{eqnarray}}
\def\eea{\end{eqnarray}}

\begin{document}
\title{Beyond Einstein's General Relativity}

\author{Francisco S. N. Lobo}

\address{Centro de Astronomia e Astrof\'{\i}sica da Universidade de Lisboa, Campo Grande, 
\\Edif\'{\i}cio C8, 1749-016 Lisboa, Portugal}

\address{Instituto de Astrof\'{\i}sica e Ci\^{e}ncias do Espa\c{c}o, Universidade de Lisboa, OAL, 
\\Tapada da Ajuda, PT1349-018 Lisboa, Portugal.}


\ead{fslobo@fc.ul.pt}

\begin{abstract}

Modern astrophysical and cosmological models are plagued with two severe theoretical difficulties, namely, the dark energy and the dark matter problems. 
Relative to the former, high-precision observational data have confirmed with startling evidence that the Universe is undergoing a phase of accelerated expansion. This phase, one of the most important and challenging current problems in cosmology, represents a new imbalance in the governing gravitational equations. Several candidates, responsible for this expansion, have been proposed in the literature, in particular, dark energy models and modified gravity, amongst others. Outstanding questions are related to the nature of this so-called ``dark energy'' that is driving the acceleration of the universe, and whether it is due to the vacuum energy or a dynamical field. On the other hand, the late-time cosmic acceleration may be due to modifications of General Relativity, which introduce new degrees of freedom to the gravitational sector itself. 
We analyze some of the modified theories of gravity that address these intriguing and exciting problems facing modern physics, and explore the foundations of gravitation theory, essential for the construction  of modified theories of gravity.

\end{abstract}

\section{Introduction}

\subsection{Need for new gravitational physics?}

A central theme in Cosmology is the perplexing fact that the Universe is currently undergoing an accelerating expansion \cite{A01}. In this context, during the last two decades Cosmology has evolved from being mainly a theoretical area of Physics to become a field supported by high precision observational data. Recent experiments call upon state of the art technology in Astronomy and in Astrophysics to provide detailed information on the contents and history of the Universe, which has led to the measuring of the parameters that describe our Universe with increasing precision. The standard model of cosmology is remarkably successful in accounting for the observed features of the Universe. However, there remain a number of fundamental open questions at the foundation of the standard model. In particular, we lack a fundamental understanding of the acceleration of the late universe. What is the so-called ``dark energy'' that is driving the acceleration of the universe? Is it a vacuum energy or a dynamical field? Or is the acceleration due to infra-red modifications of Einstein's theory of General Relativity (GR)? How is structure formation affected in these alternative scenarios? What will the outcome be of this acceleration for the future fate of the universe?
Indeed, the fact that GR is facing so many challenges, namely, the difficulty in explaining particular observations, the incompatibility with other well established theories and the lack of uniqueness, may be indicative of a need for new gravitational physics.

The resolution of these fundamental questions is extremely important for theoretical cosmology, looking beyond the standard theory of gravity. The standard model of cosmology has favoured the dark energy models as fundamental candidates responsible for the cosmic accelerated expansion. However, it is clear that these questions involve not only gravity, but also particle physics. String theory provides a synthesis of these two branches of physics and is widely believed to be moving towards a viable quantum gravity theory. One of the key predictions of string theory is the existence of extra spatial dimensions. In the brane-world scenario, motivated by recent developments in string theory, the observed 3-dimensional universe is embedded in a higher-dimensional spacetime \cite{A02}. The new degrees of freedom belong to the gravitational sector, and can be responsible for the late-time cosmic acceleration \cite{A03, A04}. 
Of course, cosmology is also an ideal testing ground for GR (in particular, the problem of the late-time cosmic acceleration). Thus, a promising approach is to assume that at large scales GR breaks down, and a more general action describes the gravitational field.

\subsection{Foundations of gravitation theory}

GR is a classical theory, therefore no reference to an action is required \cite{Sotiriou:2007yd}. Consider the Hilbert-Einstein action
\begin{equation}
S=\int d^4x \,\sqrt{-g} \left[ \frac{R}{2\kappa^2} + L_m(g^{\mu\nu}, \psi)  \right] \,,
\end{equation}    
where $R$ is the curvature scalar and $L_m(g^{\mu\nu}, \psi)$ is the matter Lagrangian, in which matter is minimally coupled to the metric $g_{\mu\nu}$ and $\psi$ collectively denotes the matter fields. However, the Lagrangian formulation is elegant, and has merits. First, at the quantum level, the action acquires a physical meaning, and a more fundamental theory of gravity will provide an effective gravitational action at a suitable limit. Secondly, it is easier to compare alternative gravitational theories through their actions rather than by their field equations. Thirdly, in many cases one has a better grasp of the physics as described through the action, by considering the couplings, kinetic and the dynamical terms.

In fact, Einstein, in deriving GR was not motivated to account for unexplained experimental results, but was driven by theoretical criteria of elegance and simplicity \cite{Will}. His primary goal was to produce a gravitation theory that incorporated the principle of equivalence and special relativity in a natural way. However, at the end of the day the theory had to be confronted with experiment, in particular, the ``three classical tests'': (i) Accounts for the perihelion advance of Mercury; (ii) Eddington's measurement of light deflection, in 1919; (iii) Pound and Rebka measure the gravitational redshift (in 1960).

It is interesting to note that in the late 1950s, Schiff and Dicke suggested that the gravitational redshift was not a true test of GR. It was purely a consequence of the equivalence principle, and did not test the field equations of gravitational theory. Note that by 1960, one may consider that the validity of GR rested upon the following empirical foundation \cite{Will} (i) One test of moderate precision (the perihelion shift of Mercury; precision: approx. 1\%); (ii) One test of low precision (the deflection of light; precision: approx. 50\%); (iii) One inconclusive test that was not a real test anyway (the gravitational redshift). In the meantime, alternative theories of gravitation were being constructed (Poincar\'{e}, Whitehead, Milne, Birkhoff, Belinfante, etc), that also laid claim to the viability of the above tests.
This triggered the development of powerful tools for distinguishing and testing theories, such as the Parametrized Post-Newtonian (PPN) expansion, pioneered by Nordvedt and extended by Nordvedt and Will \cite{Will}. Indeed, the idea that experiments test principles and not specific theories, implies the need of exploring the conceptual basis of a gravitational theory.


Probably the most unbiased assumptions to start with, in developing a gravitation theory is the Dicke framework \cite{Will}, which states that: (i) spacetime is a 4-dim manifold, with each point in the manifold corresponding to a physical event (note that a metric and affine connection is not necessary at this stage); (ii) the equations of gravity and the mathematical entities in them are to be expressed in a form that is independent of the coordinates used, i.e., in a covariant form.
It is common to think of GR, or any other gravitation theory, as a set of field equations (or an action). However, a complete and coherent axiomatic formulation of GR, or any other gravitation theory, is still lacking. One needs to formulate ``principles'' and an important one is the covariance principal, present in the Dicke Framework \cite{Sotiriou:2007yd}.

    
Thus, the basic criteria for the viability of a Gravitation Theory, following C. Will \cite{Will}, are the following. (i) It must be complete: The theory should be able to analyse from ``first principles'' the outcome of any experiment;
(ii) It must be self-consistent: Predictions should be unique and independent of the calculation method;
(iii) It must be relativistic: The theory should reduce to Special Relativity when gravity is ``turned off''; (iv) It must have the correct Newtonian limit: In the limit of weak gravitational fields and slow motion it should reproduce Newton's laws.


Note also that the Einstein Equivalence Principle (EEP) is at the heart of gravitation theory \cite{Will}. Thus if the EEP is valid, then gravitation must be a curved spacetime phenomenon, i.e., it must obey the postulates of Metric Theories of Gravity: (i) Spacetime is endowed with a metric (second rank non-degenerate tensor); (ii) The world lines of test bodies are geodesics of that metric; (iii) In local freely falling frames, Lorentz frames, the non-gravitational laws of physics are those of Special Relativity.

\subsection{Discriminating between dark energy and modified gravity models}

On the other hand, generalizations of the Einstein-Hilbert Lagrangian, including quadratic Lagrangians, involving second order curvature invariants, such as $R^{2}$, $R_{\mu \nu }R^{\mu \nu }$, $R_{\alpha\beta \mu \nu }R^{\alpha \beta \mu \nu }$, $\varepsilon ^{\alpha\beta \mu \nu }R_{\alpha \beta \gamma \delta }R_{\mu \nu }^{\gamma\delta }$, $C_{\alpha \beta \mu \nu }C^{{\alpha \beta \mu \nu}}$, have also been extensively explored \cite{A05, A06}. The main physical motivations for these modifications of gravity consist of a possibility of a more realistic representation of the gravitational fields near curvature singularities, and to create some first order approximation for the quantum theory of gravitational fields.
While these modified theories of gravity, or dark gravity, offer an alternative explanation to the standard cosmological model for the expansion history of the universe \cite{A07}, it offers a paradigm for nature fundamentally distinct from dark energy models of cosmic acceleration \cite{A08}, even those that perfectly mimic the same expansion history. It is fundamental to understand how one may differentiate these modified theories of gravity from dark energy models.

An explicit distinction between models of ``dark energy'' and ``modified gravity'' is necessary. This is due to the fact that models consisting of modifications to the gravitational sector can be mapped into a specific subset of scalar-tensor theories, namely, to a class of interacting scalar field dark energy models universally coupled to the matter species. Thus, due to this ambiguity, a practical classification is needed. In fact, taking into account solely the expansion rate of the Universe, one cannot discriminate between the dark energy models and modified gravity. However, these two alternative models will affect structure formation differently. Therefore, information on the growth of structure, at different scales and redshifts, will break the degeneracy and will serve to discriminate between both models of dark energy and modified gravity. More specifically, using linear perturbation theory, these alternative models will affect differently the constraints imposed by the Einstein equations, and the second-order growth equation will be modified, consequently changing the growth factor \cite{A09}. Thus, generic modifications of the dynamics of scalar perturbations, with respect to the $\Lambda$CDM background, can be represented by the introduction of two new degrees of freedom in 
the Einstein constraint equations, by the functions $Q(a,k)$ and $\eta(a,k)$, where $a$ is the scale factor and $k$ the perturbation scale.

In the context of modified gravity, the function $Q(a,k)$ results from a mass-screening effect due to local modifications of gravity, and effectively modifies Newton's constant. In the context of dynamical dark energy models, the function $Q(a,k)$ incorporates additional clustering, or interaction with other fields, due to the perturbations. The function $\eta$, absent in $\Lambda$CDM, parameterizes the effective stresses due to the modification of gravity or specific dynamical dark energy models. Finally, the scale and time-dependence of both functions, $Q$ and $\eta$, can be derived in the specific model considered and projected on a $(Q,\eta)$ plane. Following \cite{A09}, we denote the term ``modified gravity'' when there are additional contributions to the Poisson equation, which induces $Q \neq 1$, and where extra effective stresses arise, implying $\eta \neq 1$. Thus, ``modified gravity'' denotes models in which modifications are present in the gravitational sector and in which dark energy clusters or interacts with other fields. Thus, following this practical classification outlined in \cite{A09}, in the context of first order perturbation theory, models with $Q=\eta =1$ are denoted standard dark energy models, such as, a minimally-coupled scalar field with standard kinetic energy \cite{A09}. On the other hand, models for $Q \neq 1$ and $\eta \neq 1$ are denoted ``modified gravity'', such as scalar-tensor theories, $f(R)$ gravity \cite{A05, A06}, the DGP model \cite{A03} and generalizations \cite{A04}, massive gravity and generalized galileons \cite{A10}, Horndeski interactions \cite{A11}, bi-(multi-) gravity \cite{A12}, etc. Thus, in the context of the EUCLID mission \cite{A09}, the definitions of the functions $Q$ and $\eta$ are extremely convenient, for instance, EUCLID can distinguish between standard dynamical dark energy and modified gravity by forecasting the errors on $Q$ and $\eta$, and several combinations of these functions, such as $Q/\eta$.

\subsection{$f(R)$ gravity and extensions}

In modified gravity, in particular, in $f(R)$ gravity, one may tackle the problem using the metric formalism, which consists in varying the action with respect to the metric, although other alternative approaches have been considered in the literature, namely, the Palatini formalism \cite{A13}, where the metric and the connections are treated as separate variables; and the metric-affine formalism, where the matter part of the action now depends and is varied with respect to the connection \cite{A14}. Recently, a novel approach to modified theories of gravity that consists of adding to the Einstein-Hilbert Lagrangian an $f(R)$ term constructed a la Palatini, was proposed \cite{S10,L20}. It was shown that the theory can pass the Solar System observational constraints even if the scalar field is very light. This implies the existence of a long-range scalar field, which is able to modify the cosmological and galactic dynamics, but leaves the Solar System unaffected. These explicit models are consistent with local tests and lead to the late-time cosmic acceleration, and also verify the absence of instabilities in perturbations. This hybrid metric-Palatini theory will be briefly outlined and explored below.

The cosmological applications of generalized $f(R)$-type gravity models by assuming that the gravitational Lagrangian is given by an arbitrary function of the Ricci scalar $R$ and of the matter Lagrangian $L_m$ has also been extensively analysed \cite{S01,L38}. Specific models were explored in detail, and the gravitational field equations in the metric formalism, in the presence of a nonminimal coupling were obtained, as well as the equations of motion for test particles, which follow from the covariant divergence of the energy-momentum tensor \cite{S01, L38}. Generally, the motion is non-geodesic, and takes place in the presence of an extra force orthogonal to the four-velocity. The Newtonian limit of the equation of motion was also considered, and a procedure for obtaining the energy-momentum tensor of the matter was presented. On the other hand, the gravitational field equations are equivalent to the Einstein equations of the $f(R)$ model in empty spacetime, but differ from them, as well as from standard general relativity, in the presence of matter. Therefore the predictions of the model could lead to some major differences, as compared to the predictions of standard general relativity, or its extensions ignoring the role of matter, in several problems of current interest, such as cosmology, gravitational collapse or the generation of gravitational waves. The study of these phenomena may also provide some specific signatures and effects, which could distinguish and discriminate between the various theories of modified gravity.

\subsection{Horndeski Lagrangian}

In the context of scalar fields, these are popular building blocks used to construct models of present-day cosmological acceleration. They are appealing because such fields are ubiquitous in theories of high energy physics beyond the standard model, and, in particular, are present in theories which include extra spatial dimensions, such as those derived from string theories. The initial simplest model consisted of a minimally coupled single scalar field self-interacting through a scalar potential, and with canonical kinetic terms. This was named a quintessence field \cite{E01}. An interesting model described as coupled quintessence analysed the consequences of coupling the field responsible for acceleration to matter fields \cite{E02}. The model was rapidly extended to consider non-canonical kinetic terms, with higher powers of the field's velocity, denoted $k$-essence \cite{E03}, and non-minimal couplings to gravity. Such models included scalar tensor theories and Gauss-Bonnet dark energy. A Lagrangian based on the Galileon symmetry, with second order derivative terms, has been shown to provide a working model with interesting cosmological implications \cite{E04,E05}. Given the large number of models, the question arises how we should study and compare them in a unified manner, and determine which if any is the origin of cosmic acceleration. 

A particularly useful tool in this direction is the realisation that all these classes of models are special cases of the most general Lagrangian which leads to second order field equations. This Lagrangian was first written down by Horndeski in 1974 
\cite{A11} and rediscovered by Deffayet {\it et al} in 2011 \cite{E06,E07}. It enables researchers to adopt a unifying framework, and to determine the regions within this general theory that have appealing theoretical properties. In combination with the need to fit observations such properties are helpful in preferring regions of this general theory, and hence particular models. One example of such an appealing theoretical property is the possibility that terms within the Horndeski Lagrangian can be used to partially explain the huge discrepancy between the value of the vacuum energy in particle physics, and the value of the cosmological constant as inferred from cosmological observations. In particular these terms have been investigated with the aim of finding a viable self-tuning mechanism, which screens the spacetime curvature from the net cosmological constant \cite{E08,E09,E10}. While not yet a full theory of cosmological evolution, such a mechanism is intriguing, and leads to the question of whether realistic cosmologies can be constructed which include this mechanism as well as the late time acceleration.

\subsection{Massive gravity}

Another gravitational modification that has recently attracted much interest is the massive gravity paradigm, where instead of introducing new scalar degrees of freedom such as in $f(R)$ gravity, it modifies the graviton itself. Massive gravity is a well-defined theoretical problem by its own and has important cosmological motivations, namely, if gravity is massive it will be weaker at large scales and thus one can obtain the late-time cosmic acceleration. Fierz and Pauli presented the first linear massive gravity. However, it was shown to suffer from the van Dam-Veltman-Zakharov (vDVZ) discontinuity \cite{A21}, namely the massless limit of the results do not yield the massless theory, i.e., General Relativity. The incorporation of nonlinear terms cured the problem but introduced the Boulware-Deser (BD) ghost. This fundamental problem puzzled physicists until recently, where a specific nonlinear extension of massive gravity was proposed by de Rham, Gabadadze and Tolley (dRGT), in which the BD ghost is eliminated by a secondary Hamiltonian constraint \cite{A10}. This new nonlinear massive gravity has interesting cosmological implications, for instance, it can give rise to inflation, late-time acceleration, etc \cite{A10}. However, the basic versions of this theory exhibit instabilities at the perturbative level, and thus suitable extensions are necessary. These could be anisotropic versions, $f(R)$ extensions, bi-metric generalizations, partially-massive constructions, etc. An interesting possibility consists in a graviton mass varying function of a new scalar field \cite{A22}, where the graviton mass would be effectively larger at early cosmological times, implying interesting phenomenology both at early and late times. The crucial issue is whether one can construct a massive gravity and cosmology that can be consistent as an alternative to dark energy or other models of modified gravity, and whether this theory is in agreement with high-accuracy cosmological data, such as the growth-index or the tensor-to-scalar ratio ones, remains to be explored in detail.

\subsection{Modified teleparralel gravity and extensions}

In modified gravitational theories one usually generalizes the Einstein-Hilbert action of General Relativity, namely, one starts from the curvature description of gravity. However, a different and interesting class of modified gravity arises when one generalizes the action of the equivalent formulation of GR based on torsion. As is well known, Einstein also constructed the ``Teleparallel Equivalent of General Relativity'' (TEGR) in which the gravitational field is described by the torsion tensor and not by the curvature. In TEGR 
the corresponding Lagrangian is given by the torsion scalar $\cal{T}$, and it results from contractions of the torsion tensor, in a similar way that the Einstein-Hilbert Lagrangian $R$ results from contractions of the curvature (Riemann) tensor. Thus, in analogy with $f(R)$ gravity, one can start from TEGR and construct the $f(\cal{T})$ modified theories of gravity, by extending $\cal{T}$ to an arbitrary function in the Lagrangian \cite{A15}. The interesting feature is that although TEGR is completely equivalent to General Relativity at the level of the field equations, $f(\cal{T})$ differs radically from $f(R)$ gravity, in that they form different gravitational modifications. Hence, $f(\cal{T})$ gravity has novel and interesting cosmological implications, capable of describing inflation, the late-time acceleration \cite{A16}, large scale structure, bouncing solutions, etc. There are many open issues in $f(\cal{T})$ gravity and cosmology. In particular, one may extend this class of theories by allowing for non-minimal couplings to matter, such as $f(L_m,\cal{T})$ \cite{A17} and $f(T,\cal{T})$ \cite{A18} theories, where here $T$ is the trace of stress-energy tensor, and study their constraints from local tests and cosmology \cite{A19}, focusing on the perturbations \cite{A20}. Additionally, one may further explore the theoretical issues of the extra degrees of freedom of the theory and the different vierbein choices.

\subsection{Outline of the paper}

In this work, we briefly consider the generalized curvature-matter couplings in Section \ref{extfR}, by analysing interesting aspects and the phenomenology of the theory. In Section 
\ref{hybridmP}, we present the recent and novel approach to modified theories of gravity that consists of adding to the Einstein-Hilbert Lagrangian an $f(R)$ term constructed {\it a la} Palatini. We show that the theory can satisfy the Solar System observational constraints even if the scalar field is very light, which is consistent with cosmological scales, in particular, with the late-time cosmic acceleration. 
In Section \ref{DMhybridmP}, we consider the generalized virial theorem in the scalar-tensor representation of the hybrid metric-Palatini gravity, which can be an efficient tool in observationally testing the viability of this class of generalized gravity models.
     
\section{$f(R)$ gravity and extensions}\label{extfR}

\subsection{$f(R)$ gravity}

For instance, consider $f(R)$ gravity, for simplicity, which has an appealing feature, namely, it combines mathematical simplicity and a fair amount of generality \cite{A05}. The action is given by 
\begin{equation}
S=\int d^4x \,\sqrt{-g} \left[ \frac{f(R)}{2\kappa^2} + L_m(g^{\mu\nu}, \psi)  \right] \,,
\end{equation}  
where $f(R)$ is an arbitrary function of the Ricci scalar $R$. Varying the action with respect to the metric $g_{\mu \nu }$ yields the field equations, given by
\begin{equation}
F(R)R_{\mu \nu }-\frac{1}{2}f(R)g_{\mu \nu }-\nabla_\mu
\nabla_\nu \,F(R)+g_{\mu\nu} \nabla^\alpha
\nabla_\alpha F(R)
= \kappa^2  T_{\mu \nu }^{(m)}~,
\label{field}
\end{equation}
where we have denoted $F(R)=f'(R)$, and the prime represents the derivative with respect to the scalar curvature. 
The matter energy-momentum tensor is defined as
\begin{equation}
T_{\mu \nu
}^{(m)}=-\frac{2}{\sqrt{-g}}\frac{\delta(\sqrt{-g}\,{\cal
L}_m)}{\delta(g^{\mu\nu})} ~.
\end{equation}    

Note that the Ricci scalar is a dynamical degree of freedom, which is transparent from the trace equation given by $F R -2f + 3 \nabla_\mu \nabla^\mu F = \kappa T$ (where $F=df/dR$). This introduces a new light scalar degree of freedom, which consequently produces a late-time cosmic acceleration. However, the light scalar strongly violates the Solar System constraints and the way out of this problem is through the `chameleon' mechanism, i.e., the scalar field becomes massive in the Solar System. The approaches to $f(R)$ gravity consist in the metric, Palatini, and metric-affine formalisms (and one may mention a fourth approach, i.e., the hybrid metric-Palatini formalism which will be explored below, in some detail).

\subsection{Geons in quadratic Palatini gravity}

We mention an interesting application in the Palatini formalism, by considering the action \cite{Lobo:2013adx}:
\begin{eqnarray}\label{eq:action}
S[g,\Gamma,\psi_m]=\frac{1}{2\kappa^2}\int d^4x \sqrt{-g}\left[R+l_P^2 \left(a R^2+R_{\mu\nu}R^{\mu\nu} \right)\right]
-\frac{1}{16\pi}\int d^4x \sqrt{-g}F_{\mu\nu}F^{\mu\nu} \ ,
\end{eqnarray}
where $\kappa^2\equiv 8\pi G/c^4$, $l_P^2$ represents the Planck length squared, $a$ is a free parameter,  $F_{\mu\nu}$  is the electromagnetic field strength, $g_{\mu\nu}$ is the space-time metric, $R=g^{\mu\nu}R_{\mu\nu}$, $R_{\mu\nu}={R^\rho}_{\mu\rho\nu}=R_{\nu\mu}$, and
${R^\alpha}_{\beta\mu\nu}=\partial_{\mu}
\Gamma^{\alpha}_{\nu\beta}-\partial_{\nu}
\Gamma^{\alpha}_{\mu\beta}+\Gamma^{\alpha}_{\mu\lambda}\Gamma^{\lambda}_{\nu\beta}-\Gamma^{\alpha}_{\nu\lambda}\Gamma^{\lambda}_{\mu\beta} $. The connection $\Gamma^{\alpha}_{\mu\nu}$ is {\it a priori} independent of the metric (Palatini formalism) and must be determined by the field equations.

More specifically, it was found that the end state of black hole evaporation can be represented by non-singular and without event horizon stable solitonic remnants with masses of the order the Planck scale. Though these objects are locally indistinguishable from spherically symmetric, massive electric (or magnetic) charges, they turn out to be sourceless geons containing a wormhole generated by the electromagnetic field. The results were obtained by interpreting semiclassical corrections to Einstein's theory in the first-order (Palatini) formalism, which yields second-order equations and avoids the instabilities of the usual (metric) formulation of quadratic gravity.

This construction has interesting applications. For instance, it has recently been suggested that Einstein-Rosen (ER) bridges can be interpreted as maximally entangled states of two black holes that form a complex Einstein-Podolsky-Rosen (EPR) pair \cite{Maldacena:2013xja}, in the context of the firewall debate \cite{Almheiri:2012rt, Almheiri:2013hfa} . This relationship has been dubbed as the $ER = EPR$ correlation. The latter conjecture was also considered in the context of the quadratic Palatini theory \cite{Lobo:2014fma}. An important result, which stems from the underlying assumptions about the geometry on which the theory is constructed, is the fact that all the charged solutions of the quadratic Palatini theory possess a wormhole structure. The results show that spacetime may have a foam-like microstructure with wormholes generated by fluctuations of the quantum vacuum. This involves the spontaneous creation/annihilation of entangled particle-antiparticle pairs, existing in a maximally entangled state connected by a non-traversable wormhole. Since the particles are produced from the vacuum and therefore exist in a singlet state, they are necessarily entangled with one another, which provides further support to the $ER=EPR$ claim.

Furthermore, the collapse of a charged radiation fluid in a Planck-suppressed quadratic extension of GR was also considered \cite{Lobo:2013vga}. Exact analytical solutions were obtained that extend the charged Vaidya-type solution of GR, which allows to explore in detail new physics at the Planck scale. Starting from Minkowski space, it was found that the collapsing fluid generates wormholes supported by the electric field. The relevance of the findings in relation to the quantum foam structure of space-time and the meaning of curvature divergences in this theory was also discussed. The above analysis was extended in \cite{Lobo:2014zla}, where the physical significance of curvature divergences in this theory and the topology change issue was discussed. This supports the view that space-time could have a foam-like microstructure pervaded by wormholes generated by quantum gravitational effects.

\subsection{Nonminimal curvature-matter coupling}

$f(R)$ modified theories of gravity can be generalized by introducing in the action a linear nonminimal coupling between matter and geometry, given by the following action
\begin{equation}
S=\int \left\{\frac{1}{2}f_1(R)+\left[1+\lambda f_2(R)\right]{
L}_{m}\right\} \sqrt{-g}\;d^{4}x~,
\end{equation}
where $f_i(R)$ (with $i=1,2$) are arbitrary functions of the Ricci
scalar $R$ and ${L}_{m}$ is the matter Lagrangian density \cite{S01} .
    
Varying the action with respect to the metric $g_{\mu \nu }$
yields the field equations, given by
\begin{eqnarray}
F_1(R)R_{\mu \nu }-\frac{1}{2}f_1(R)g_{\mu \nu }-\nabla_\mu \nabla_\nu \,F_1(R)+g_{\mu\nu} \nabla_\alpha \nabla^\alpha F_1(R)
=-2\lambda F_2(R){L}_m R_{\mu\nu}
   \nonumber \\
+2\lambda(\nabla_\mu
\nabla_\nu-g_{\mu\nu}\nabla_\alpha \nabla^\alpha){L}_m F_2(R)
+[1+\lambda f_2(R)]T_{\mu \nu }^{(m)}~,
\label{field1b}
\end{eqnarray}
where we have denoted $F_i(R)=f'_i(R)$, and the prime represents
the derivative with respect to the scalar curvature.

One verifies the non-conservation of the energy-momentum tensor
\begin{equation}
\nabla^\mu T_{\mu \nu }^{(m)}=\frac{\lambda F_2}{1+\lambda
f_2}\left[g_{\mu\nu}{\cal L}_m- T_{\mu \nu
}^{(m)}\right]\nabla^\mu R ~. \label{cons1}
\end{equation}
Thus, the coupling between the matter and the higher derivative curvature terms describes an exchange of energy and momentum between both. Analogous couplings arise after a conformal transformation in scalar-tensor theories of gravity (and string theory). In the absence of the coupling, one verifies the conservation of the energy-momentum, which can also be verified from the diffeomorphism invariance of the matter part of the action. 

In order to test the motion in the model, consider a perfect fluid, so that the
equation of motion for a fluid element is given by
\begin{equation}
\frac{Du^{\mu }}{ds} \equiv \frac{du^{\mu }}{ds}+\Gamma _{\alpha
\beta }^{\mu }u^{\alpha }u^{\beta }=f^{\mu }~, \label{eq1}
\end{equation}
where $f^\mu$ is an extra-force given by
\begin{eqnarray}
\label{force}
f^{\mu }&=&\frac{1}{\rho +p}\Bigg[\frac{\lambda
F_2}{1+\lambda f_2}\left({L}_m-p\right)\nabla_\nu
R+\nabla_\nu p \Bigg] h^{\mu \nu }\,.
\end{eqnarray}
with $h_{\mu \lambda }=g_{\mu \lambda }+u_{\mu }u_{\lambda }$ the projection operator. The extra force $f^{\mu}$ is orthogonal to the four-velocity of the particle, $f^{\mu }u_{\mu }=0$.

A particularly intriguing feature is that the extra force depends on the form of the Lagrangian density. Note that considering the Lagrangian density $L_m = p$, where $p$ is the pressure, the extra-force vanishes \cite{Bertolami:2008ab}. However, it has been argued that this is not the unique choice for the matter Lagrangian density and that more natural forms for $L_m$, such as $L_m = -\rho$, do not imply the vanishing of the extra-force. Indeed, in the presence of the nonminimal coupling, they give rise to two distinct theories with different predictions \cite{Faraoni:2009rk}.
We refer the reader to \cite{Bertolami:2008zh,Harko:2014gwa} for a more detailed discussion.

\subsection{Generalized curvature-matter couplings}

\subsubsection{$f(R,L_m)$ gravity:}

One may further generalize the linear curvature-matter coupling by assuming that the gravitational Lagrangian is given by an arbitrary function of the Ricci scalar $R$ and of the matter Lagrangian $L_m$ \cite{Harko:2010mv}. This essentially consists in a maximal extension of the Hilbert–Einstein action, and the action takes the following form
\begin{equation}
S=\int f\left(R,L_m\right) \sqrt{-g}\;d^{4}x\,.
   \label{action}
\end{equation}

Varying the action with respect to the metric, yields the following field equation
\begin{eqnarray}\label{field2a}
&&f_{R}\left( R,L_{m}\right) R_{\mu \nu }+\left( g_{\mu \nu } 
\nabla
_{\alpha}\nabla _{\alpha} -\nabla
_{\mu }\nabla _{\nu }\right) f_{R}\left( R,L_{m}\right)
\nonumber\\
&&-\frac{1}{2}\left[
f\left( R,L_{m}\right) -f_{L_{m}}\left( R,L_{m}\right)L_{m}\right] g_{\mu \nu }=
\frac{1}{2}%
f_{L_{m}}\left( R,L_{m}\right) T_{\mu \nu }.
\end{eqnarray}
Considering $f\left( R,L_{m}\right) =R/2+L_{m}$, the Hilbert-Einstein Lagrangian, we
recover the standard Einstein field equation of general relativity,  $R_{\mu \nu
}-(1/2)g_{\mu \nu }R=T_{\mu \nu }$.

This theory possess extremely interesting properties, such as the
covariant divergence of the energy-momentum tensor is non-zero
\begin{eqnarray}
\nabla ^{\mu }T_{\mu \nu }=2\nabla
^{\mu }\ln \left[ f_{L_m}\left(R,L_m\right) \right] \frac{\partial L_{m}}{
\partial g^{\mu \nu }}\,.  \label{noncons}
\end{eqnarray}
The requirement of $\nabla ^{\mu }T_{\mu \nu }=0$ yields an effective
functional relation between the matter Lagrangian density and the
function $f_{L_m}\left(R,L_m\right)$, given by 
\begin{equation}
\nabla ^{\mu }\ln
\left[ f_{L_m}\left(R,L_m\right) \right] \partial L_{m}/
\partial g^{\mu \nu }=0 \,.
\end{equation}
Note  that considering a specific matter Lagrangian density, by an appropriate choice of the function $f(R,L_m)$, one can construct, at least in principle, conservative models with arbitrary curvature-matter dependence. We refer the reader to \cite{Harko:2014gwa} for more details.

\subsubsection{$f(R,T)$ gravity:}

Consider another extension of GR, namely, $f (R,T)$ gravity, where the gravitational Lagrangian is given by an arbitrary function of the Ricci scalar $R$ and of the trace of the energy-momentum tensor $T$ \cite{Harko:2011kv}. The action takes the following form
\begin{equation}
S=\frac{1}{2\kappa^2}\int
f\left(R,T\right)\sqrt{-g}\;d^{4}x+\int{L_{m}\sqrt{-g}\;d^{4}x}\,.
\end{equation}
Note that the dependence from $T$ may be induced by exotic imperfect fluids or quantum effects (conformal anomaly). This theory may be considered a relativistically covariant model of interacting dark energy. The $f(R,T)$ gravitational model mentioned above has been given a great amount of recent attention and we refer the reader to \cite{Harko:2014gwa,Harko:2011kv} for more details.

\subsubsection{$f(R,T,R_{\mu\nu}T^{\mu\nu})$ gravity:}

The $f(R,T)$ gravitational theory \cite{Harko:2011kv} for the specific case of a traceless energy-momentum tensor, $T=0$, for instance, when the electromagnetic field is involved, reduces to $f(R)$ gravity and all non-minimal couplings of gravity to the matter field vanish. This fact  motivated a further generalization of $f(R,T)$ gravity that consists in including an explicit first order coupling between the matter energy-momentum $T_{\mu \nu}$ and the Ricci tensor \cite{Haghani:2013oma, Odintsov:2013iba}.
It is interesting to note that in contrast to $f(R,T)$ gravity, for $T=0$, this extra coupling still has a non-minimal coupling to the electromagnetic field via the $R_{\mu\nu}T^{\mu\nu}$ coupling term in the action, which is non-zero in general.

The action, is given by
\begin{equation}\label{eq200}
S=\frac{1}{2\kappa^2}\int d^4x\sqrt{-g}f\left(R,T,R_{\mu\nu}T^{\mu\nu}\right)+\int
d^4x\sqrt{-g}L_m \,.
\end{equation}
The only requirement imposed on the function $f\left(R,T,R_{\mu\nu}T^{\mu\nu}\right)$ is that it  is an arbitrary analytical function in all arguments. We refer the reader to \cite{Haghani:2013oma, Odintsov:2013iba} for more details.

\section{Hybrid metric-Palatini gravity}\label{hybridmP}

Recently, a novel approach to modified theories of gravity that consists of adding to the Einstein-Hilbert Lagrangian an $f(R)$ term constructed {\it a la} Palatini, was proposed \cite{S10}. It was shown that the theory can satisfy the Solar System observational constraints even if the scalar field is very light. This implies the existence of a long-range scalar field, which is able to modify the cosmological and galactic dynamics, but leaves the Solar System unaffected. These explicit models are consistent with local tests and lead to the late-time cosmic acceleration, and also verify the absence of instabilities in perturbations. Criteria to obtain cosmic acceleration were discussed and the field equations were also analyzed as a dynamical system \cite{L20}. In this context, several classes of dynamical cosmological solutions, depending on the functional form of the effective scalar field potential, describing both accelerating and decelerating Universes are explicitly obtained. Furthermore, the cosmological perturbation equations were derived and applied to uncover the nature of the propagating scalar degree of freedom and the signatures these models predict in the large-scale structure \cite{L20}. Thus, this novel model proves to have extremely interesting properties, and deserves a further in depth study. 

The action of the hybrid metric-Palatini theory is given by \cite{S10,L20}
\begin{equation} \label{eq:S_hybrid}
S= \frac{1}{2\kappa^2}\int d^4 x \sqrt{-g} \left[ R + f(\R)\right] +S_m \ ,
\end{equation}
where $S_m$ is the matter action, $\kappa^2\equiv 8\pi G$, $R$ is the Einstein-Hilbert term, 
$\R \equiv  g^{\mu\nu}\R_{\mu\nu} $ is the Palatini curvature, and $\R_{\mu\nu}$ is defined in 
terms of an independent connection $\hat{\Gamma}^\alpha_{\mu\nu}$  as
$\R_{\mu\nu} \equiv \hat{\Gamma}^\alpha_{\mu\nu , \alpha} - \hat{\Gamma}^\alpha_{\mu\alpha , 
\nu} + \hat{\Gamma}^\alpha_{\alpha\lambda}\hat{\Gamma}^\lambda_{\mu\nu} 
-\hat{\Gamma}^\alpha_{\mu\lambda}\hat{\Gamma}^\lambda_{\alpha\nu}$.

The action (\ref{eq:S_hybrid}) may be expressed as the following scalar-tensor theory \cite{S10}
\begin{equation} \label{eq:S_scalar2}
S=\int \frac{d^4 x \sqrt{-g} }{2\kappa^2}\left[ (1+\phi)R +\frac{3}{2\phi}\partial_\mu \phi
\partial^\mu \phi -V(\phi)\right]+S_m .
\end{equation}
This action differs from the $w=-3/2$ Brans-Dicke theory in the coupling of the scalar to the curvature, which in the $w=-3/2$ theory is of the form $\phi R$. This simple modification will have important physical consequences. 

\subsection{Weak-field, slow-motion behaviour}

Taking into account the effects of the scalar field $\phi$ on the solar system dynamics by studying the weak-field limit of the field equations, we consider an expansion of the metric and the scalar field about a cosmological solution. This sets the asymptotic boundary values, using a quasi-Minkowskian coordinate system, form which one deduces the effective Newton constant $G_{\rm eff}$ and the post-Newtonian parameter (PPN) $\gamma$, which are given by
\begin{eqnarray}\label{g_eff}
G_{\rm eff} \equiv  \frac{G}{1+\phi_0}\left[1-\left(\phi_0/3\right)e^{-m_\varphi r}\right]\,, 
  \qquad
\gamma  \equiv  \frac{1+\left(\phi_0/3\right)e^{-m_\varphi r}}{1-\left(\phi_0/3\right)e^{-
m_\varphi r}} \,,
\end{eqnarray}
respectively (we refer the reader to \cite{S10} for specific details). As is clear from the above expressions, the coupling of the scalar field to the local system 
depends on the amplitude of the background value $\phi_0$. If $\phi_0$ is small, then $G_{\rm 
eff}\approx G$ and $\gamma\approx 1$ regardless of the value of the effective mass 
$m_\varphi^2$.

This contrasts with the result obtained in the metric version of $f(R)$ 
theories:
\begin{eqnarray}
G_{\rm eff} \equiv   G
\left(1+e^{-m_f r}/3\right)/\phi_0   \,,
\qquad
\gamma \equiv   \left(1-\frac{e^{-m_f r}}{3}\right)\Big/\left(1+\frac{e^{-m_f r}}{3}\right) \,,
\end{eqnarray}
which requires a large mass $m_f^2\equiv (\phi
V_{\phi\phi}-V_\phi)/3$ to make the Yukawa-type corrections negligible in local experiments.

\subsection{Late-time cosmic speedup}

As a specific example of modified cosmological dynamics, consider the spatially flat Friedman-Robertson- Walker (FRW) metric, $ds^2=-dt^2+a^2(t) d{\bf x}^2$. The modified Friedmann equations are given by
\begin{eqnarray}
3H^2&=& \frac{1}{1+\phi }\left[\kappa^2\rho +\frac{V}{2}-3\dot{\phi}\left(H+\frac{\dot{\phi}}
{4\phi}\right)\right] \ ,\label{field1} \\
2\dot{H}&=&\frac{1}{1+\phi }\left[ -\kappa^2(\rho+P)+H\dot{\phi}+\frac{3}
{2}\frac{\dot{\phi}^2}{\phi}-\ddot{\phi}\right] \,, \label{field2}
\end{eqnarray}
respectively, and the scalar field equation is
\begin{equation}
\ddot{\phi}+3H\dot{\phi}-\frac{\dot{\phi}^2}{2\phi}+\frac{\phi}{3}
[2V-(1+\phi)V_\phi]=-\frac{\phi\kappa^2}{3}(\rho-3P) \ .  \label{3}
\end{equation}

The qualitative behavior of the scalar field can be read directly from the above by 
rewriting it as follows
\begin{equation}
\ddot{\phi}+3H\dot{\phi}-\frac{\dot{\phi}^2}{2\phi}+M^2_\phi(T)\phi=0 \ ,  \label{3a}
\end{equation}
where $T=-(\rho-3P)$ and we have defined $M^2_\phi(T)$ as
\begin{equation}\label{eq:mass}
M^2_\phi(T)\equiv m_\phi^2-\frac{1}{3}\kappa^2T=\frac{1}{3}[2V-(1+\phi)V_\phi-\kappa^2T] \ ,
\end{equation}


We now propose a model that is consistent at Solar System and cosmological scales, and which is constructed on grounds of mathematical simplicity. Consider for mathematical simplicity:
\begin{equation} \label{pot1}
V(\phi)=V_0+V_1\phi^2\,.
\end{equation}
The trace of the field equation automatically implies $R=-\kappa^2T+2V_0$. As $T\to 0$ with the cosmic expansion, the solution naturally evolves into a de Sitter phase ($V_0\sim \Lambda$) for consistency with observations. If $V_1$ is positive, the de Sitter regime represents the minimum of the potential.  Note that the effective mass for local experiments, $m_\varphi^2=2(V_0-2 V_1 \phi)/3$, is positive if $\phi<V_0/V_1$. For $V_1 \gg V_0 $, the amplitude is small enough to pass Solar System tests. Thus, the exact de Sitter solution is compatible with dynamics of the scalar field in this model.

\section{Dark matter in modified gravity and the generalized virial theorem}\label{DMhybridmP}


In the context of dark matter, two observations, namely, the behaviour of the galactic rotation 
curves and the mass discrepancy in galactic clusters, suggest the existence of a (non or weakly 
interacting) form of dark matter at galactic and extra-galactic scales. The gravitational masses of clusters of galaxies are estimated by assuming a hydrostatic equilibrium of both the hot intra-cluster gas and of the galaxies with the binding cluster potential. Therefore, the total mass of a cluster of galaxies can be estimated in two ways: First, by taking into account the dynamical motions of the member galaxies of the cluster, and with the application of the virial theorem,  one obtains an estimate $M_{V}$ for the mass of the cluster. Second, the total baryonic mass $M_B$ can be determined  by adding the mass of each individual galaxy member of the cluster. The mass discrepancy at the galactic cluster level arises as observations show that $M_{V}$ is much greater than $M_B$, with typical values of $M_{V}/M_B \sim 20-30$.

Thus, generally, the dynamics of test particles around galaxies, as well as the corresponding mass deficit, is explained by postulating the existence of a hypothetical dark matter. In fact, the behavior of the rotation curves shows the existence of a constant velocity region, near the baryonic matter distribution, followed by a quick decay at large distances. In \cite{Capozziello:2013yha}, the possibility was considered that the behavior of the rotational velocities of test particles gravitating around galaxies can be explained within the framework of the hybrid metric-Palatini gravitational theory. We refer the reader to related work in other modified theories of gravity \cite{effdark}.

In the intermediate galactic scale, it was shown explicitly that in the hybrid metric-Palatini model the tangential velocity can be explicitly obtained as a function of the scalar field of the equivalent scalar-tensor description. The possibility of constraining the form of the scalar field and the parameters of the model by using the stellar velocity dispersions was also analyzed. All the physical and geometrical quantities and the numerical parameters in the hybrid metric-Palatini model can be expressed in terms of observable/measurable parameters, such as the tangential velocity, the baryonic mass of the galaxy, the Doppler frequency shifts, and the stellar dispersion velocity, respectively. Therefore, the results obtained open the possibility of testing the hybrid metric-Palatini gravitational models at the galactic or extra-galactic scale by using direct astronomical and astrophysical observations.

Here, we consider the generalized virial theorem in the scalar-tensor representation of the hybrid metric-Palatini gravity, which can be an efficient tool in observationally testing the viability of this class of generalized gravity models \cite{Capozziello:2012qt}.

\subsection{Generalized virial theorem in hybrid metric-Palatini gravity}

Consider a self-gravitating system of identical, collisionless point particles in random motion.
To obtain the basic field equations we will use the scalar-tensor representation of hybrid metric-Palatini gravity, given by action (\ref{eq:S_scalar2}), which allows a clear physical interpretation of the model.

The metric of an isolated  spherically symmetric
cluster is given by
\be
ds^{2}=-e^{\nu \left( r\right) }dt^{2}+e^{\lambda \left( r\right)
}dr^{2}+r^{2}\left( d\theta ^{2}+\sin^{2}\theta d\varphi
^{2}\right).  \label{line}
\ee
The galaxies in the cluster are considered identical and collisionless point
particles, and their space-time distribution  is described by a distribution function $f_B$.

The distribution function $f_B$ obeys the general relativistic Boltzmann equation, which is the transport equation for the distribution function for a system of particles in a curved arbitrary Riemannian space-time.
The general relativistic Boltzmann equation is provided by
\be  \label{distr}
\left( p^{\alpha }\frac{\partial }{\partial x^{\alpha }}-p^{\alpha }p^{\beta
}\Gamma _{\alpha \beta }^{i}\frac{\partial }{\partial p^{i}}\right) f_B=0,
\ee
where $p^{\alpha }$ is the four-momentum of the particle, and $\Gamma _{\alpha \beta }^{i}$ are the Christoffel symbols associated to the metric. 
Note that the collissionless Boltzmann equation states that the local phase space density viewed by an observer co-moving with a star or galaxy is conserved.

Thus, the energy-momentum tensor of matter is given by
\be
T_{\mu \nu }=\int f_B\, m\, u_{\mu }u_{\nu }\;du,
\ee
where $m$ is the mass of the particle (galaxy), $u_{\mu }=\left(u_t,u_r,u_{\theta},u_{\varphi}\right)$ is the
four-velocity of the galaxy, with $u_t$ denoting the temporal component, and $du=du_{r}du_{\theta}du_{\varphi}/u_{t}$ is the invariant volume element of the velocity space.

The energy-momentum tensor $T_{\mu \nu }$ can be represented in terms of an effective density $\rho_{\mathrm{eff}}$ and of two effective anisotropic pressures, the radial $p_{\mathrm{eff}}^{(r)}$ and the
tangential $p_{\mathrm{eff}}^{(\perp)}$ pressures, respectively, given by
\be
\rho_{\mathrm{eff}} = \rho \left\langle u_{t}^{2}\right\rangle, p_{%
\mathrm{eff}}^{(r)}=\rho \left\langle u_{r}^{2}\right\rangle,
p_{\mathrm{eff}}^{(\perp)} = \rho \left\langle u_{\theta
}^{2}\right\rangle= \rho \left\langle u_{\varphi }^{2}\right\rangle,
\ee
where $\rho $ is the mass density of the ordinary baryonic matter, and $\left\langle u_{i}^{2}\right\rangle $, $i=t,r,\theta ,\varphi $ is the average
value of $u_{i}^{2}$, $i=t,r,\theta ,\varphi $.

By using this form of the energy-momentum tensor, the summing up the components of the gravitational field equations describing a cluster of galaxies in hybrid metric-Palatini gravity, takes the form
\be
e^{-\lambda }\left( \frac{\nu ^{\prime \prime }}{2}+\frac{\nu ^{\prime 2}}{4}
+\frac{\nu ^{\prime }}{r}-\frac{\nu ^{\prime }\lambda ^{\prime }}{4}\right)
\simeq 4\pi G\rho \left\langle u^{2}\right\rangle +4\pi G\rho _{\phi }^{(eff)} \,,
\label{ff1}
\ee
and the relativistic Boltzmann equation finally takes the form
\bea
\int_{0}^{R}4\pi \rho \left[ \left\langle u_{1}^{2}\right\rangle
+\left\langle u_{2}^{2}\right\rangle +\left\langle u_{3}^{2}\right\rangle%
\right] r^{2}dr 
-\frac{1}{2}\int_{0}^{R}4\pi r^{3}\rho \left[ \left\langle
u_{0}^{2}\right\rangle +\left\langle u_{1}^{2}\right\rangle\right] \frac{%
\partial \nu }{\partial r}dr=0.  \label{kin}
\eea
We refer the reader to \cite{Capozziello:2012qt} for explicit details.

\subsection{Geometric quantities}

Since we are interested in astrophysical applications at the extra-galactic
level, we may assume that the deviations from standard general relativity
are small, i.e., $\phi \ll 1$.
We assume that the approximations that apply to test particles in stable circular motion around galaxies, also apply to galactic clusters. First, assume $\nu $ and $\lambda $ are slowly varying functions of $r$ (i.e.~$\nu^{\prime}$ and $\lambda^{\prime}$ are small), so that all the quadratic terms are neglected.
Secondly, assume that the motion of the galaxies is non-relativistic, so that
they  have velocities much smaller than the velocity of the light, i.e., $\langle u_{1}^{2}\rangle \approx \langle u_{2}^{2}\rangle \approx
\langle u_{3}^{2}\rangle \ll \langle u_{0}^{2}\rangle \approx 1$.

Thus, the gravitational field equations reduce to  
\be  \label{fin1}
\frac{1}{2r^{2}}\frac{\partial }{\partial r}\left(r^{2} \frac{\partial \nu }{%
\partial r}\right) = 4\pi G\rho + 4\pi G\rho_{\phi}^{(eff)},
\ee
and the relativistic Boltzmann equation to
\be
2K-\frac{1}{2}\int_{0}^{R}4\pi r^{3}\rho \frac{\partial \nu }{\partial r}%
dr=0,  \label{cond1}
\ee
respectively, where
\begin{equation}
K=\int_{0}^{R}2\pi \rho \left[ \left\langle u_{1}^{2}\right\rangle
+\left\langle u_{2}^{2}\right\rangle +\left\langle u_{3}^{2}\right\rangle %
\right] r^{2}dr,
\end{equation}
is the total kinetic energy of the galaxies. 

Consider that the gravitational potential energies of the cluster are defined as:
\begin{eqnarray}
\Omega _B=-\int_{0}^{R}\frac{GM_B(r)}{r}\,dM_B(r), \qquad
\Omega _{\phi }^{(eff)}=\int_{0}^{R}\frac{GM_{\phi }^{(eff)}(r)}{r}\,dM_B(r),
\end{eqnarray}
respectively, where $R$ is the cluster radius.
Since in hybrid metric-Palatini gravity, the quantity $M_{\phi }^{(eff)}$ has essentially a geometric origin, we tentatively denote it as the \textit{geometric mass} of the
cluster, defined as
\begin{equation}  \label{darkmass}
M_{\phi }^{(eff)}\left( r\right) =4\pi \int_{0}^{r}\rho _{\phi}^{(eff)}(r')r'^{2}
dr'.
\end{equation}

Using the above equations, we arrive at the generalization of
the virial theorem, which takes the familiar
form
\be
2K + \Omega = 0  \label{theor} \,,
\ee
where $\Omega $ is total gravitational potential energy of the system defined by
$\Omega = \Omega _B- \Omega_{\phi }^{(eff)} $,
which contains a term consisting of a geometric origin, $\Omega_{\phi }^{(eff)} $.

The generalized virial theorem, can be represented  in a more transparent physical  form if we introduce the radii $R_{V}$ and $R_{\phi }$, defined by
\bea
R_{V}=M_B^{2}\Bigl/\int_{0}^{R}\frac{M_B(r)}{r}\,dM_B(r),\Bigr.  \qquad
R_{\phi }^{(eff)}=\left[M_{\phi }^{(eff)}\right]^{2}\Bigl/\int_{0}^{R}\frac{M_{\phi }^{(eff)}(r)}{r}\,dM_B(r),\Bigr.
\label{RU3}
\eea
respectively.  
In analogy to the geometric mass considered above, the quantity $R_{\phi }$ may be denoted as the \textit{geometric radius} of the cluster of galaxies.

Thus, the baryonic potential energy  $\Omega _B$ and the effective scalar field potential energy $\Omega _{\phi}^{(eff)}$ are finally given by
\begin{eqnarray}
\Omega _B =-\frac{GM_B^{2}}{R_{V}}, \qquad
\Omega _{\phi}^{(eff)} =\frac{G\left[M_{\phi }^{(eff)}\right]^{2}}{R_{\phi }^{(eff)}},
\end{eqnarray}
respectively.

Now, we define the virial mass $M_{V}$ of the cluster of galaxies as
\begin{equation}
2K=\frac{GM_BM_{V}}{R_{V}}.
\end{equation}
After substitution into the virial theorem, we obtain the following relation between the virial and the baryonic mass of the galaxy cluster
\be  \label{fin6}
\frac{M_{V}}{M_B}=1+\frac{\left[M_{\phi }^{(eff)}\right]^{2}R_{V}}{M_B^{2}R_{\phi }^{(eff)}}.
\ee

If $M_{V}/M_B>3$, a condition which holds for most of the observed galactic
clusters, then Eq.~(\ref{fin6}) provides the virial mass in hybrid metric-Palatini gravity,
which can be approximated as
\be
M_{V}\approx \frac{\left[M_{\phi}^{(eff)}\right]^2}{M_B}\frac{R_{V}}{R_{\phi }^{(eff)}}.  \label{virial}
\ee

In the present model there is also a strict proportionality between the virial mass of the cluster and its baryonic mass, a relation which can also be tested observationally. 
From the point of view of the astrophysical observations the virial mass $M_V$ is determined from the
study of the velocity dispersion $\sigma _r^2$ of the stars and of the galaxies in the clusters. 
According to the virial theorem in hybrid metric-Palatini gravity, most of the mass in a cluster with mass $M_{tot}$ is in the form of the geometric mass $M_{\phi }^{(eff)}$, so that $M_{\phi}^{(eff)} \approx M_{tot}$.

Thus, since galaxy clusters are ``dark'' matter
dominated
objects, the main contribution to their mass comes from the geometric mass $%
M_{\phi }$, so that with a very good approximation we have $M_{\phi }\approx
M_{V}\approx M_{tot} $. 
Therefore the virial theorem immediately provides the
following mass scaling relation
\begin{equation}
M_{V}\approx M\frac{R_{\phi }}{R_{V}}.
\end{equation}
This equation shows that the virial mass is proportional to the baryonic
(normal) mass of the cluster, and that the ratio of the total mass and of
baryonic mass is determined by a purely geometric quantity, the geometric
radius $R_{\phi }$. 

Hence the geometric radius of the cluster can be
determined from observations, once the virial and baryonic masses and the
virial radius, respectively, are known.

\section{Conclusions}

While dark gravity offers an alternative explanation to the standard cosmological model for the expansion history of the universe, it offers a paradigm for nature fundamentally distinct from dark energy models of cosmic acceleration, even those that perfectly mimic the same expansion history. It is fundamental to understand how one may differentiate these modified theories of gravity from dark energy models. All modified gravity models induce observational signatures at the post-Newtonian level, which are translated by the parameterized post-Newtonian (PPN) metric coefficients arising from these extensions of GR. Tests from the solar system, large scale structure and lensing, as well as laboratory and space-based Equivalence Principle experiments, essentially restrict the range of allowed modified gravity models.
 
Surveys such as the EUCLID space telescope, the Square Kilometre Array (SKA) radio telescope, the Dark Energy Survey (DES), and the Extended Baryon Oscillation Spectroscopic Survey (eBOSS) as part of the Sloan Digital Sky Survey III (SDSS)  will provide new opportunities to test the different cosmological models.
Indeed, with the wealth of unprecedented high precision observational data that will become available by these upcoming and planned surveys, we are dawning in a golden age of cosmology, which offers a window into understanding the perplexing nature of the cosmic acceleration, dark matter and of gravity itself.

\section*{Acknowledgements}
I would like to deeply thank all my collaborators and co-athours related to the work presented in this paper.  
I am supported by a Funda\c{c}\~{a}o para a Ci\^{e}ncia e Tecnologia Investigador FCT Research contract, with reference IF/00859/2012, funded by FCT/MCTES (Portugal), and acknowledge financial support of the Funda\c{c}\~{a}o para a Ci\^{e}ncia e Tecnologia through the grant EXPL/FIS-AST/1608/2013. 


\end{document}